\documentclass[floats,aps,twocolumn,showpacs]{revtex4}
\usepackage{dcolumn}

\usepackage{epsfig}
\usepackage{longtable}
\newcommand {\la} {\langle}
\newcommand {\ra} {\rangle}
\newcommand {\beq} {\begin{eqnarray}}
\newcommand {\eeqn} [1] {\label{#1} \end{eqnarray}}%

\newcommand {\ve} [1] {\mbox{\boldmath $#1$}}

\begin{document}
%
%

\title{
Asymptotic normalization coefficients for mirror virtual nucleon decays
in a microscopic cluster model }

\author{
N.\ K.\ Timofeyuk$^{1)}$ and P. Descouvemont$^{2)}$
}                 

\affiliation{
$^{1)}$ Department of Physics, 
University of Surrey, Guildford,
Surrey GU2 7XH, England, UK\\
$^{2)}$ Physique Nucl\'{e}aire Th\'{e}orique et Physique 
Math\'{e}matique, CP229\\
Universit\'{e} Libre de Bruxelles, B1050 Brussels, Belgium
}


\date{\today}

\begin{abstract}
It has been suggested  recently ($Phys. Rev. Lett.$ 91, 232501 (2003))
that charge symmetry of nucleon-nucleon interactions
 relates the Asymptotic Normalization Coefficients (ANCs)
of proton and neutron virtual  decays of mirror nuclei. 
This relation is given by a simple analytical
formula which  involves  
 proton and neutron separation energies, charges of residual nuclei and
the range of their strong interaction with the  last nucleon.  
Relation between mirror ANCs, if understood properly, 
can be used to predict astrophysically relevant direct proton capture cross
sections  using neutron ANCs measured with stable beams.
In this work, we calculate one-nucleon ANCs for several   light
mirror pairs, using  microscopic
two-, three- and four-cluster
models, and compare the  ratio of mirror ANCs to the predictions of
the simple analytic formula. We also investigate mirror
symmetry between other characteristics of mirror 
one-nucleon overlap integrals,  namely, spectroscopic factors  
and single-particle ANCs.

\end{abstract}

\maketitle

\section{Introduction}

The asymptotic normalization coefficient (ANC) for
one-nucleon virtual decay $A \rightarrow (A-1) + N$ is one 
of the fundamental characteristics of a nucleus $A$.
It determines the magnitude of the large 
distance behaviour of the projection of the  bound state
wave function of the nucleus $A$ onto the binary 
channel $(A-1)+N$. The recent interest in studying the
one-nucleon ANCs is due to the role which they play in
nuclear astrophysics for predicting cross sections
of non-resonant capture reactions at stellar energies. 
The ANCs provide overall normalization of the astrophysical
$S$-factors of such reactions.
Since the same ANCs play a crucial role in other 
peripheral processes such as  transfer 
reactions, they can be measured in  laboratories
and used to predict non-resonant capture processes at
low stellar  energies \cite{xu}.

To determine  relevant to astrophysics proton ANCs from transfer reactions, 
the use of  radioactive beams is often required, 
which generally involves more difficult and less 
accurate experiments than those possible with stable beams.
At the same time,  stable beams can often be used to determine 
neutron ANCs associated with mirror virtual one-neutron decays.
This has been noticed some time ago
in Refs. \cite{Tim,TI}, where the one-nucleon ANCs of the
mirror pairs $^8$B $-$ $^8$Li
and $^{12}$N $-$ $^{12}$B were studied 
in a microscopic approach. In these  works, the calculated ANCs themselves
depended  strongly on the choice of the nucleon-nucleon (NN) force but the 
ratios of ANCs for mirror pairs were  practically
independent of the choice of the NN force. 
This property of the ANC ratios could be used to predict  proton capture
rates at astrophysical energies from   information about  mirror ANC's obtained 
from transfer reactions with stable beams. 
A first experiment which uses the  idea of Ref. \cite{Tim} to deduce
the ANC of $^8$B from the $^8$Li ANC 
has been already performed \cite{Tra03}.

Recently, it has been  pointed out that the ANCs
for mirror virtual decays $^AZ_N \rightarrow \,^{A-1}Z_{N-1}+n $
and $^AN_Z \rightarrow \,^{A-1}N_{Z-1}+p$ are related  if the charge
symmetry of nucleon-nucleon (NN) interactions is satisfied \cite{Tim03}.
This link is approximated  by a simple analytical formula which is a
consequence of the relation between the on-shell amplitudes of
mirror virtual decays. These on-shell amplitudes, called vertex constants,
are equivalent to
the coupling constants in particle physics \cite{BBD77}.

A link between  mirror ANCs  also follows from the single-particle model
of nuclei if  charge symmetry is valid
both for single-particle potential wells and for mirror one-nucleon 
spectroscopic factors.
As shown in  Ref. \cite{Tim03},  predictions of
such a single-particle model are close to the predictions of
the simple analytical formula, derived from  consideration
of mirror vertex constants, if nucleon separation energies are 
relatively large.
This agreement deteriorates with decrease of separation
energies and for weakly-bound $s$-states with nodes,
the difference between the two different estimates
for the ratio of mirror ANCs can reach  $\sim 15-20\%$. 
 
At present, more accurate but simple approximations  relating mirror ANCs
are not available. Therefore, 
numerical calculations
using theoretical structure models are very important.
In the present paper, we try to improve our understanding of
relation between mirror ANCs by performing  calculations
 within a microscopic cluster
model (MCM).  This model considers the  many-body nature of
atomic nuclei and takes into account differences in
nuclear structure arising because of charge symmetry breaking
due to the Coulomb interaction.
We expect that, in MCM, the lack of accuracy of  the
two different
approximations  from Ref. \cite{Tim03} is reduced.
We calculate one-nucleon overlap integrals for
some  mirror  light nuclei and
concentrate  mainly on  mirror ANCs, but 
other properties of overlap  integrals, such as spectroscopic
factors, r.m.s. radii and single-particle ANCs, and their mirror
symmetry are   investigated.

In Sec. II we give  definitions  for  ANCs, 
their expressions via nuclear wave functions,
show the approximations
for the ratio  of mirror ANCs, derived in Ref. \cite{Tim03}, 
and discuss their validity. In Sec. III we briefly describe
our microscopic cluster model and the ANCs associated with it.
The results obtained in microscopic calculations are discussed in Sec. IV.
A summary and conclusions are presented in Sec. V.


\section{Overlap integrals and ANCs for mirror virtual decays}

The ANC $C_{lj}$ for the one-nucleon virtual decay $A \rightarrow B + N$,
where $B = A -1$, 
is defined  via the tail of  the overlap integral $I_{lj}(r)$
\beq
 I_{lj}(r) = \la
 \chi_{\frac{1}{2}\tau} [[ Y_{l}(\hat{\ve{r}}) \otimes \chi_{\frac{1}{2}} ]_j
\otimes \Psi^{J_B}]_{J_A}|\Psi^{J_A}\ra
\eeqn{oi}
between the many-body wave functions $\Psi^{J_A}$ and $\Psi^{J_B}$ 
of nuclei $A$ and $B$.
Here $l$ 
is the  orbital momentum, $j$ is the total relative angular momentum
between $B$ and $N$, $\tau$ is the isospin projection and
$ \chi_{\frac{1}{2}\tau} $ is the isospin wave function of 
nucleon $N$, and $r$ is the distance 
between $N$  and the center-of-mass of $B$. 
Asymptotically, this  overlap behaves as
\beq
\sqrt{A}\,I_{lj}(r)\approx C_{lj}\frac{W_{-\eta,l+1/2}(2\kappa r)}
{r}, 
\,\,\,\,\,\, 
r\rightarrow\infty,
\eeqn{anc}
where  
$\kappa =(2\mu\epsilon/\hbar^2)^{1/2}$, $\epsilon$ is the one-nucleon
separation energy, $\eta=Z_BZ_Ne^2\mu/\hbar^2\kappa $, 
$\mu$ is the reduced mass for the $B + N$ system and $W$ is the Whittaker 
function. According to Ref. \cite{BBD77}, the ANC $C_{lj}$, multiplied
by the trivial factor $i^l \pi^{\frac{1}{2}} (\hbar/\mu c)$, is
equal to the on-shell amplitude (or vertex constant) of the
one-nucleon virtual decay  $A \rightarrow B + N$. This vertex constant
can be written as a matrix element that contains
the many-body wave functions of the nuclei $A$ 
and $B$. Therefore,
the ANC $C_{lj}$ can  also be represented by the same
matrix element as follows \cite{MT90,Tim} \footnote{We have found a mistake
in the phase factor  of the function  $\varphi_l(i\kappa r)$
in Refs. \cite{MT90, Tim, Tim03}. The corrected phase factor is given
in   Eq. (\ref{phi}) of the present work. 
This mistake does not influence the  published previously  
ANCs, spectroscopic factors or shapes of overlap  integrals.}
(\ref{phi}) is:
\beq
 C_{lj}  &=& -\frac{2\mu \sqrt{A}}{\hbar^2} \nonumber \\&\times & 
\la \chi_{\frac{1}{2}\tau} [[ \varphi_l(i\kappa r)
Y_{l}(\hat{\ve{r}}) \otimes \chi_{\frac{1}{2}} ]_j
\otimes \Psi^{J_B}]_{J_A}||\hat{{\cal V}}||\Psi^{J_A}\ra,
\eeqn{vff}
where 
\beq
\varphi_l(i\kappa r) = 
 e^{-\frac{\pi i}{2}(l+1+\eta) } F_l(i\kappa r)/{\kappa r}, 
\eeqn{phi}
$F_l$ is the regular Coulomb wave function at imaginary momentum
$i\kappa$,  and
\beq
\hat{{\cal V}} = 
\sum_{i=1}^B V_{NN} (|\ve{r}_i-\ve{r}_A|)+\Delta V_{Coul} =
\hat{{\cal V}}_N+\Delta V_{Coul},
\eeqn{V}
\beq
\Delta V_{Coul} = \sum_{i=1}^B 
\frac{e_ie_A} {|\ve{r}_i-\ve{r}_A|} -\frac{Z_Be_Ae}{r}. 
\eeqn{dV}
Here $e_i$ ($e_A$) is the charge of the $i$-th ($A$-th)
nucleon, $Z_B$ is the charge of the residual nucleus $B$ 
and $V_{NN}$ is the two-body nuclear NN potential.
If the separated nucleon is  a neutron,
$\varphi_l(i\kappa r) = i^{-l} j_l(i\kappa r)$ and
$j_l(i\kappa r)$ is   the spherical Bessel function.

It has been shown in Ref. \cite{Tim03}
that the ratio 
\beq
{\cal R}= \left(\frac{C_p}{C_n}\right)^2,
\eeqn{ratio}
where $C_p$ and $C_n$ are
proton and neutron  ANCs for mirror nucleon decays, can be approximated
 as follows:
\beq
{\cal R } \approx {\cal R}_0 \equiv \left|\frac{F_l(i\kappa_pR_N)}
{\kappa_pR_N\,j_l(i\kappa_nR_N)}\right|^2.
\eeqn{rn}
Here $\kappa_p$  and $\kappa_n$ are determined by the 
proton and neutron separation  energies $\epsilon_p$ and $\epsilon_n$
and $R_N$ is the
radius of the nuclear interior to the choice of which 
the ratio ${\cal R}_0$ is not strongly sensitive. 

The approximation (\ref{rn})
has been derived in Ref. \cite{Tim03} using Eq. (\ref{vff}) for mirror
decays and assuming that
1) non-monopole on $r$ contributions from  $\Delta V_{Coul}$  are negligible;
2) differences in mirror wave functions inside the nuclear interior
due
to Coulomb interaction are not important
and
3) charge symmetry of strong interactions is valid.
As  has been mentioned in Ref. \cite{Tim03}, non-monopole contributions from
 $\Delta V_{Coul}$  
increase  the ratio  ${\cal R}$. 
On the other hand,  due to the stronger Coulomb interactions in  $Z > N$ nuclei
the magnitude of their wave functions  
are smaller in the nuclear interior as compared to the wave functions
of $Z < N$ nuclei. This should lead  to decrease of  ${\cal R}$, which 
may become 
 more noticeable for very small
proton separation energy.  Besides, if any nodes are present in the
overlap $I_{lj}(r)$ then the   contributions  from $r > R_N$
to $C_{lj}$,
determined by Eq. (\ref{vff}),  may become larger. 
This can introduce further uncertainties into
approximation  (\ref{rn}) because   
differences in mirror proton and neutron wave functions in the $r > R_N$ region
are important due to the Coulomb
effects.
It is possible, however, that all different factors may compensate each 
other so that,
finally, the approximation (\ref{rn}) could be accurate enough to be used 
in practical
purposes in the absence of more advanced  detailed calculations.

Another approximation 
for ${\cal R}$ can be obtained 
if the overlap integral $I_{lj}(r)$ is thought of
as being a normalised single-particle wave function times  
the spectroscopic factor $S$.
In this case $C_{p(n)} = \sqrt{S_{p(n)}}\, b_{p(n)}$, where $b_{p(n)}$ 
is the single-particle proton (neutron)  ANC. If charge symmetry is assumed
both for the mirror single-particle wells and the mirror spectroscopic factors,
then the ratio ${\cal R}$ is equal to the single-particle ratio
\beq
{\cal R} \approx {\cal R}_{s.p.} \equiv (b_p^{c.s.}/b_n^{c.s.})^2,
\eeqn{rsp} 
where $b_p^{c.s.}$ and $b_n^{c.s.}$ are calculated for exactly the same
nuclear potential well.
The accuracy of the approximation    (\ref{rsp}) 
is determined by the following factors:
(i)
the  two-body potential model does not include
effects of long-range contributions 
from non-monopole terms in $\Delta V_{Coul}$; 
(ii)
the single particle potential wells for mirror pairs may differ
because of slightly different matter distributions in their cores and 
(iii) the spectroscopic
factors for mirror pairs may be not exactly the same.

Below, to understand better the validity of these approximations,
we perform  calculations of ${\cal R}$
for some light nuclei
based on a microscopic cluster model.


\section{One-nucleon ANCs in a  microscopic cluster model}

The cluster wave function for a nucleus $A$ consisting of 
a core $B$ and a nucleon $N$ can be represented as follows:
\beq
\Psi^{J_AM_A} = \sum_{lSJ_B \omega} 
{\cal A} [  \chi_{\frac{1}{2}\tau} [  g_{\omega lS}^{J_B} (\ve{r}) \otimes 
[\Psi^{J_B}_{\omega}\otimes  \chi_{\frac{1}{2}}
 ]_S]_{J_A M_A} ] 
\eeqn{9}
where ${\cal A} = A^{-\frac{1}{2}}(1-\sum_{i=1}^{A-1} P_{i,A})$ and
the operator $P_{i,A}$ permutes spatial and spin-isospin coordinates
of the $i$-th and $A$-th nucleons.
In this work,
$\Psi^{J_B }_{\omega}$ is a wave function of nucleus $B$ 
with the angular momentum $J_B$
defined 
either in  translation-invariant harmonic-oscillator shell model, either 
in a multicluster model. 
The quantum number $\omega$ labels  
states with the same angular momentum $J_B$ and $S$ is the channel spin. 
The relative wave function $g_{\omega lS}^{J_B} (\ve{r}) = 
g_{\omega lS}^{J_B}(r) \,Y_{lm}(\hat{r})$ also depends on  
$J_B$ and is determined from the solution of the Schr\"odinger
equation for $\Psi^{J_AM_A} $ with some chosen NN potential. 
Below,
we skip $J_B$ and $\omega$ in relative functions, overlap integrals
and their characteristics for simplicity of notations.

The main advantage of a microscopic cluster model (MCM) is that it is able
to provide the correct asymptotic behaviour for the overlap integral
between $A$ and $B$.
At large distances, $r \rightarrow \infty$,
where the antisymmetrization between the external nucleon 
and the core is negligible, this overlap   behaves as
\beq
I_{ lS}(r) \approx
A^{-\frac{1}{2}} g_{lS}(r) \approx A^{-\frac{1}{2}} C_{lS}
 \frac{ W_{-\eta,l+1/2}(2\kappa r)}{r}.  
\,
\eeqn{10}
We achieve this type of behaviour by using 
the microscopic R-matrix
approach \cite{BHL77} and
 determine
the ANC 
$C_{ls}$ from the asymptotic behaviour of the relative wave 
functions corresponding to the $\omega$ state components \cite{BT92}.

The MCM has been formulated in the $lS$ coupling scheme and the
transition to the $lj$ coupling scheme is given 
by the standard transformation
\beq
C_{lj} = \sum_S (-)^{J_B + \frac{1}{2} -S} 
\hat{S} \hat{j} \,  W(J_B \frac{1}{2}J_A l; Sj) \, C_{lS},
\eeqn{6}
where $W$ is the Racah coefficient and $\hat{x} = (2x+1)^{1/2}$.
The same transformation is applicable to overlap 
integrals $I_{lj}$ and $I_{lS}$.

The MCM should provide more reliable ratios
 ${\cal R}$ for mirror ANCs
than the
approximations (\ref{rn}) and (\ref{rsp}). Indeed, unlike in
Eq. (\ref{rn}),
the differences in the internal structure of mirror nuclei
due to the Coulomb interaction are taken into account in the MCM.
Also, determining the ANC directly from the tail of the overlap
means that all the non-monopole contributions
from  $\Delta V_{Coul}$ are present in the proton ANCs.
The effects of  core excitations are included as well.
On the other hand, the MCM does not appeal to the concept of single-particle
structure of nuclei  and it does not need the hypothesis about charge
symmetry for mirror single-particle potential
wells and mirror spectroscopic factors. Charge symmetry for these
quantities can still be studied within the MCM by investigating
mirror spectroscopic factors, defined as
norms of the
MCM overlap integrals:
\beq
S_{lj} = A \int_0^{\infty} dr \, r^2 (I_{lj}(r))^2,
\eeqn{sf}
and
the single-particle ANCs $b_{lj} = C_{lj}S_{lj}^{-1/2}$.
The latter is possible because
the  overlap integrals $I_{lj}(r)$, divided by the square root of their
spectroscopic factors $S_{lj}$, are normalised functions
of only one degree of freedom and they play the same role as  
single-particle wave functions generated by some effective local 
single-particle potential.
Comparison between   single-particle ANCs $b_{lj}$
for mirror nuclei may help
to understand if mirror symmetry of the effective local
single-particle potential wells 
is valid.

\section{Ratio of mirror ANCs in the MCM}

\subsection{Mirror ANCs with charge independent NN interactions}

First of all, we have calculated   ANCs for several nuclei
assuming that  NN interactions in mirror states are exactly the same.
This assumption does not allow us to simultaneously reproduce the 
experimental neutron and proton separation
energies in mirror states. However, it will enables us to explore
the validity of the approximations (\ref{rn}) and (\ref{rsp}). 
The effective NN interactions, used in this work, are 
 the  Volkov potential V2 \cite{volkov}
and the Minnesota (MN) potential \cite{minnesota}.
The two-body spin-orbit force
\cite{BP81} and the Coulomb interaction are also included.

\begin{figure*}[t]
\centerline{\psfig{figure=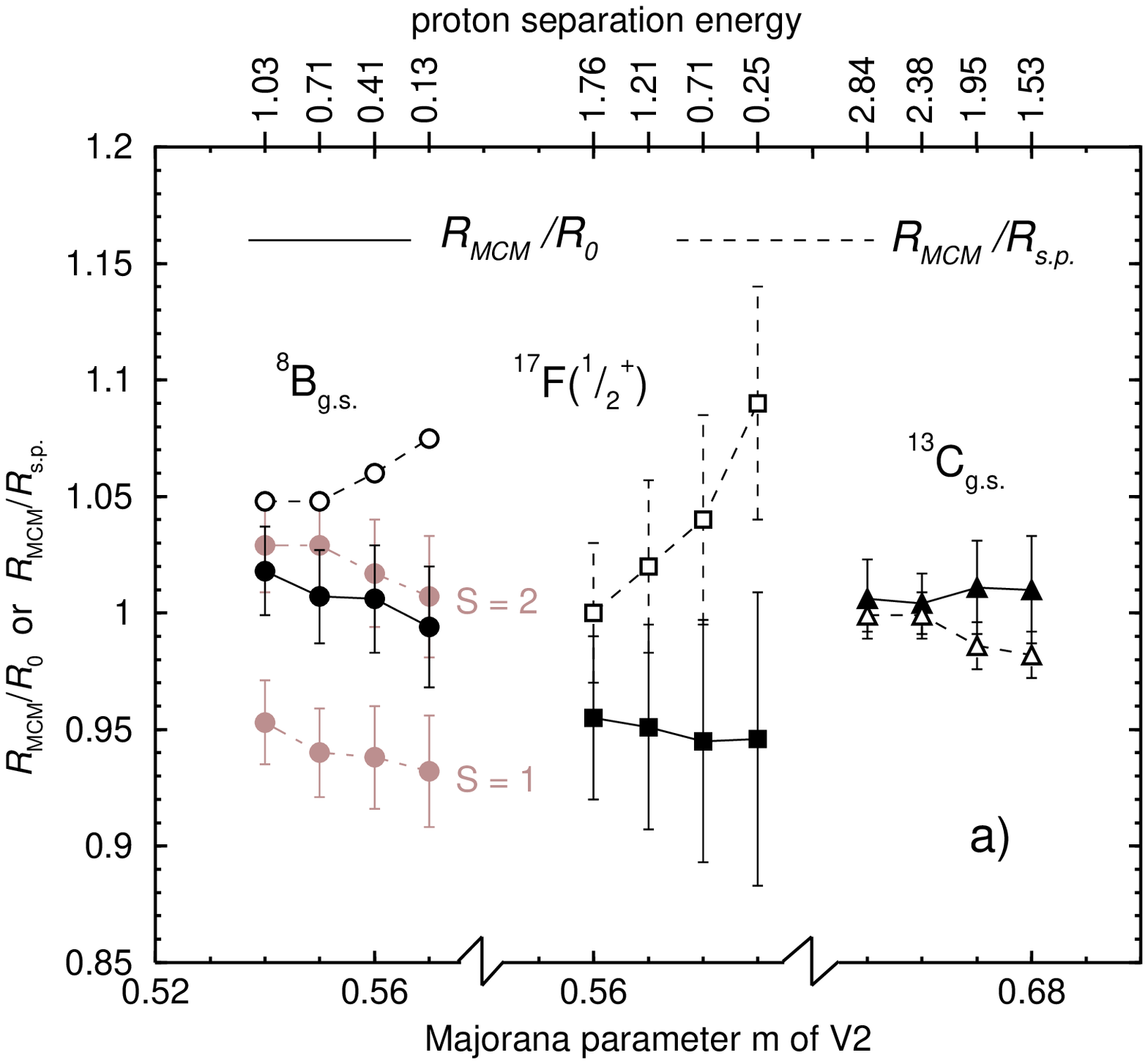,width=0.46\textwidth}
             \psfig{figure=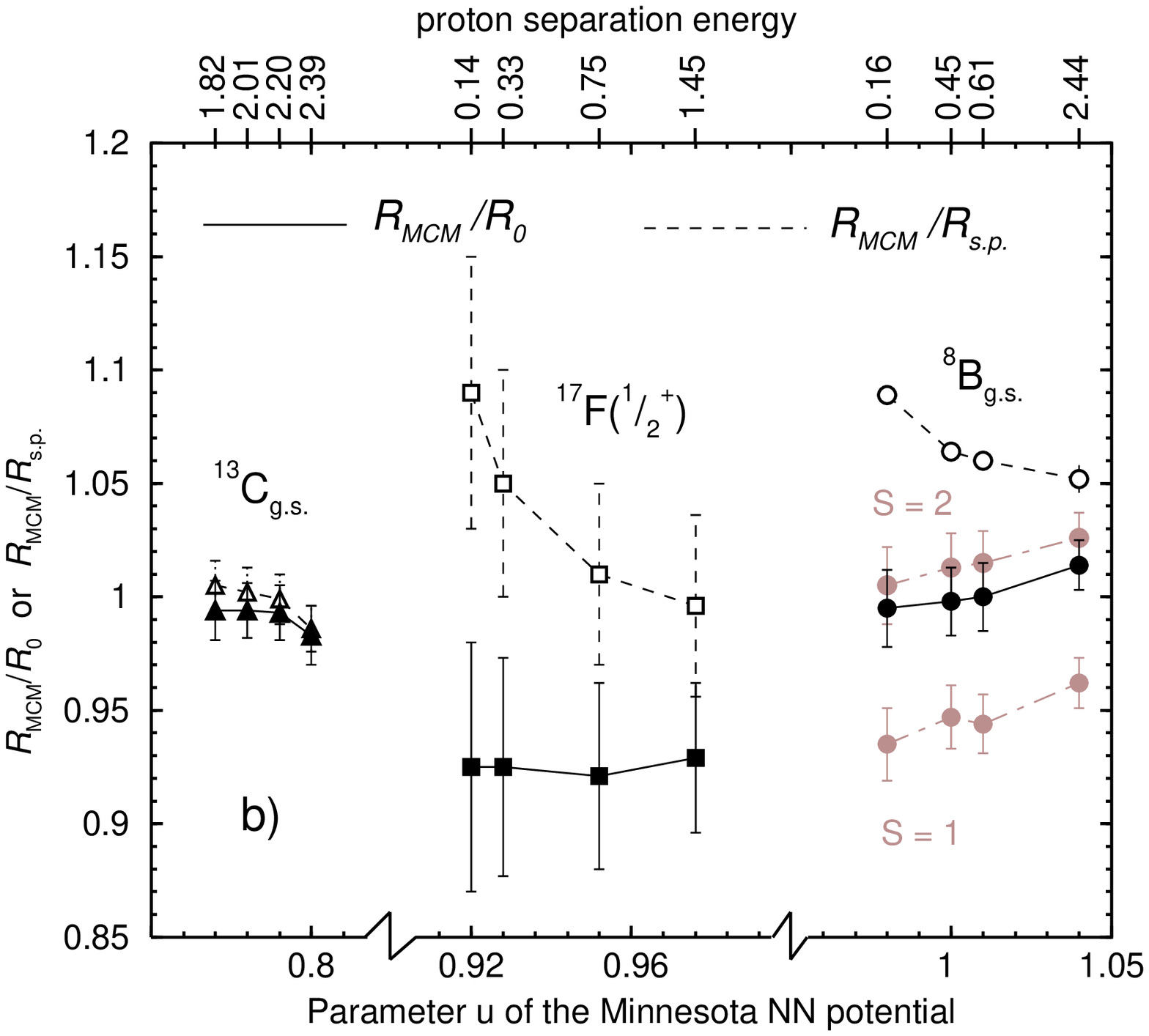,width=0.46\textwidth}}

\caption{Ratio ${\cal R}_{MCM}/{\cal R}_0$ (solid curves connecting
black symbols) and ${\cal R}_{MCM}/{\cal R}_{s.p.}$ (dashed curves
connecting open symbols) for different values of the Majorana parameter
$m$ of the Volkov potential V2 ($a$) and
and for different parameter $u$ of the MN force ($b$). The proton separation
energies corresponding to each calculation are shown above upper 
horizontal axes.
Grey symbols correspond to different
channel spin $S$ in the $^8$B-$^8$Li mirror pair.
}
\end{figure*}

In this section, 
we have considered three  
mirror pairs: $^8$B(2$^+ )- ^8$Li(2$^+$), 
$^{13}$N$(\frac{1}{2}^-) - ^{13}$C$(\frac{1}{2}^-)$ and 
$^{17}$F$(\frac{1}{2}^+) - ^{17}$O$(\frac{1}{2}^+)$,
which have been previously studied in Refs. \cite{Des04,TBD97,BT92,BDH98}
in the $\alpha + ^3$He + p  ($\alpha$ + t + n),
$^{12}$C + p  ($^{12}$C + n) and
$^{16}$O + p ($^{16}$O + n) microscopic cluster models.
We have  calculated the ANCs for these mirror pairs 
for several values of the  parameters $m$  and  $u$
of the V2 and MN interactions 
 chosen to provide a range of theoretical
separation energies  covering the
experimental separation energies. 
For each value of $m$ and $u$ we have calculated
the ratio  ${\cal R}_{MCM} = (C^{MCM}_p/C_n^{MCM})^2$, using
theoretical separation energies,
and compared it to the analytical value ${\cal R}_0$ and 
single-particle estomate ${\cal R}_{s.p.}$
given by Eqs. (\ref{rn}) and (\ref{rsp}).
The ratios
${\cal R}_{MCM}/{\cal R}_0$ and ${\cal R}_{MCM}/{\cal R}_{s.p.}$ 
are shown in Fig. 1.

The error bars on Fig. 1 reflect the following uncertainties in the
calculations of ${\cal R}_0$ and ${\cal R}_{s.p.}$. 
 ${\cal R}_0$ depends on the range $R_N$ of the interaction
potential between  last nucleon $N$ and the core $B$. In Ref. \cite{Tim03} this
range was taken as $1.3B^{1/3}$. In fact, some contributions from
the NN potential at larger $R_N$ may not be negligible, especially
for cases when the wave function of the last nucleon has nodes.
We have observed that, for all nuclei considered up to  now,  ${\cal R}_0$
slowly increases with $R_N$, reaches its maximum slightly
beyond than $1.3B^{1/3}$ and then  slowly decreases.
In estimating uncertainties in ${\cal R}_0$, we have assumed that its value
is somewhere
between $1.3B^{1/3}$ and the maximum value.
As for  ${\cal R}_{s.p.}$, its uncertainties are due to the residual dependence
on the nucleon-core potential. We have chosen this potential in the Woods-Saxon
form and have varied its depth and  radius at fixed diffusenesses
to reproduce simultaneously the theoretical proton and neutron separation
energies, calculated in the MCM. The  uncertainties   in ${\cal R}_0$ 
and ${\cal R}_{s.p.}$ vary
with the choice of a  mirror pair
and are the largest for   weakly-bound proton  $s$-states  with
a node in their wave functions.

As shown in Fig.1, the precision of ${\cal R}_0$ 
and ${\cal R}_{s.p.}$ in approximating  ${\cal R}_{MCM}$ varies for
different systems.
For the relatively strongly bound mirror pair 
$^{13}$N$- ^{13}$C, with the last nucleon
 in the $p$-wave with respect to the $^{12}$C core,  
${\cal R}_{MCM}$ agrees with ${\cal R}_0$ and ${\cal R}_{s.p.}$ 
within these uncertainties.

Another 0p-shell mirror pair, $^8$B(2$^+ )- ^8$Li(2$^+$), is significantly
less bound than $^{13}$N  - $^{13}$C. 
However, the quality of agreement 
between ${\cal R}_{MCM}$ and ${\cal R}_0$ for the
 ANCs squared summed over the channel spin, 
$C_l^2=C_{l S=1}^2+C_{l S=2}^2$,   is the same
as in the $^{13}$N - $^{13}$C case (see black solid curves in Fig.1). 
In contrast,
the single-particle estimate 
${\cal R}_{s.p.}$ is larger than  ${\cal R}_{MCM}$ 
and this difference
increases with decreasing proton separation energy reaching 9$\%$.
We recall that it is 
$C_l^2$ that determine the cross sections of the radiative capture 
reaction $^7$Be(p,$\gamma)^8$B.

The  ratios
${\cal R}_{MCM}$,  calculated for spin channels $S$=1 and
$S$=2, differ  
by  $\sim 10\%$ (see grey symbols in Fig.1). 
If the  wave functions of   mirror nuclei
were exactly the same, then the ratio of mirror ANCs   would  not depend
on the channel spin. 
The charge symmetry breaking due to the Coulomb interaction
may manifest itself  stronger
in small components of the wave functions. Therefore, for such components,
 deviation from (\ref{rn}) can be more noticeable.
 Indeed, for $^8$B(2$^+ )- ^8$Li(2$^+$),
$C_{lS=2}^2$ are about four times larger than $C_{lS=1}^2$
and  ${\cal R}_{MCM}$  for $S$=2 agrees
with ${\cal R}_0$ better than  in the channel with $S$=1.

In the last mirror pair considered in this section,
$^{17}$F$(\frac{1}{2}^+) - ^{17}$O$(\frac{1}{2}^+)$, 
the valence proton and neutron are  in  1$s$-state with respect
to the core $^{16}$O. 
The calculated ${\cal R}_{MCM}$  values are about 5 to 8$\%$ smaller than
${\cal R}_0$ for all the  proton separation
energies considered. At the same time, ${\cal R}_{MCM}$ agrees   with
single-particle ratio ${\cal R}_{s.p.}$ if the the proton separation energy
becomes larger than 1.4 MeV. 
When the NN interaction  is changed so that
the proton separation energy decrease down to 0.13 MeV then ${\cal R}_{s.p.}$
overestimates ${\cal R}_{MCM}$ by about 9$\%$.


\subsection{Mirror ANCs with charge symmetry breaking NN interactions}

Charge symmetry in realistic NN interactions is broken and this may be
reflected in
effective NN interactions. In the MCM
calculations, 
different  parameters $m$  and $u$ of the V2 and MN 
potentials should be taken in mirror states in order
to achieve   agreement between   theoretical 
and experimental separation energies. 
A different choice of $m$ and $u$ in mirror states   means that
 charge symmetry is still present
in even NN interactions, but   odd NN interactions  
are scaled with some renormalisation factor. 
We refer   to this different choice as
to charge symmetry breaking
for the sake of simplicity, however, we do realise
that it is not the same as for realistic NN potentials.

In this section, we calculate ANCs for
several mirror pairs of nuclei that  have
two-, three- or four- cluster
structure. In most cases, the wave functions of these nuclei have been
obtained earlier. We calculate ANCs  in the $lj$ coupling
scheme  as usually done in the analysis of transfer reactions,
in which these ANCs can, or have been,  determined.
For nuclear astrophysics,
the sum of the ANCs squared $C_l^2 = C_{lj=1/2}^2+C_{lj=3/2}^2$ 
is often
needed rather then their individual values in channels
with different $j$. We show these values as well.
Other characteristics of one-nucleon overlaps  $\la A|A-1 \ra$,
namely, spectroscopic factors, r.m.s. radii and single-particle ANCs
$b_{lj}^2=C_{lj}^2/S_{lj}$ are presented is this section as well.

\subsubsection{$^8{\rm B} -  ^8{\rm Li}$}

To reproduce experimental values of
both the  proton and neutron separation energies,  
 the Majorana parameters $m$ of V2 should   differ in $^8$B  and $^8$Li by
1.8$\%$. For the MN potential, this difference is only 1.0$\%$.

The   $C_{1\frac{3}{2}}^2$ values
obtained with the V2 potential
are by 22-26$\%$ larger than those calculated with MN (see Table I). 
However, the
ratio ${\cal R}_{\frac{3}{2}}= C^2_{1\frac{3}{2}}(p)/C^2_{1\frac{3}{2}}(n)$ 
changes only by 3$\%$ 
with the NN potential choice.  These ratios, 1.048 for V2 and 1.079 for MN,
are smaller than the value ${\cal R}_0$ = 1.13 $\pm$ 0.01 predicted
by the formula (\ref{rn}) but higher than the single-particle value
${\cal R}_{s.p.}$ = 1.01$\pm$0.01 obtained from equality of mirror proton 
and neutron single-particle potential
wells and the mirror proton and neutron spectroscopic factors.

 The $C_{1\frac{1}{2}}^2$ values are much smaller
then $C_{1\frac{3}{2}}^2$ and they change only by 9$\%$ for $^8$B
and 4$\%$ for $^8$Li with
different NN potential choices. The ratio ${\cal R}_{\frac{1}{2}}
= C^2_{1\frac{1}{2}}(p)/C^2_{1\frac{1}{2}}(n)$ of the mirror ANCs
in this case, 1.26 and 1.19 for the V2 and MN potentials
respectively, are by 20 $\%$ and 10$\%$ larger 
than ${\cal R}_{\frac{3}{2}}$, which should be due to the stronger influence of
charge symmetry breaking effects in the small $j = 1/2$ component.

The  $C_l^2$  value increases by 20$\%$ 
with a change of the NN force. However, the ratio ${\cal R}_{MCM}$ 
changes only within 2$\%$,
being 1.068 and 1.092 for V2 and MN respectively. Its
average value of 1.08 is closer to the analytical
value ${\cal R}_0$ = 1.13 $\pm$ 0.01   than to the single-particle
value ${\cal R}_{s.p.}$ = 1.01$\pm$0.01.
We recall that for charge independent NN interactions, 
the difference in ${\cal R}_{MCM}$
and ${\cal R}_0$ is only about 2$\%$ for energies $\epsilon_p$ 
similar to the experimental ones.

The proton ANCs for $^8$B have been determined in Ref. \cite{azh}
using the ($^7$Be,$^8$B) transfer reactions
on two different targets,
$^{14}$N and $^{10}$B.
The average value  of $C_1^2$  deduced from these
experiments is  0.449$\pm$0.045 fm$^{-1}$. 
The breakup reaction
at intermediate energies gave a very close value of 0.450$\pm$0.039  fm$^{-1}$
\cite{Tra01}.
These values are  by 42$\%$ and 30 $\%$  smaller than the MCM predictions  
with the V2 and MN potentials respectively.  
The neutron ANC of the mirror nucleus £$^8$Li has been
experimentally determined  in Ref. \cite{Livius} 
using transfer reaction $^{13}$C($^7$Li,$^8$Li)$^{12}$C. 
Its value,
$C_1^2$ = 0.449$\pm$0.045 fm$^{-1}$, is also much smaller than the
predictions of 
the MCM. However, the ratio of the experimentally
determined mirror ANCs is 1.08$\pm$0.15,
which in excellent agreement with the average MCM ratio.

The shapes of the angular distributions
of the
transfer reaction  $^{13}$C($^7$Li,$^8$Li)$^{12}$C are very sensitive
to the interference between the contributions
from the overlap integrals with different $j$. It was
found in Ref. \cite{Livius} that $C^2_{1\frac{3}{2}}/C^2_{1\frac{1}{2}}$ 
= 0.13(2).
This value is in excellent agreement
with the  value of 0.131 obtained in the MCM using the MN potential. 
As for V2,
it predicts for $C^2_{1\frac{3}{2}}/C^2_{1\frac{1}{2}}$ 
much lower ratio, equal to 0.108.

The mirror spectroscopic factors
$S_{1\frac{3}{2}}$ 
of large components of the  overlap integrals  
 differ by about 1$\%$  
and they change less than by 1$\%$ with different choice
of the NN potential (see Table I).
However, small spectroscopic factors $S_{1\frac{1}{2}}$ are
more sensitive to the NN potentials choice and their difference in mirror states
reaches   20$\%$. The sums of mirror
spectroscopic factors   $S_{1} = S_{1\frac{3}{2}} + S_{1\frac{1}{2}}$
differ in  mirror nuclei $^8$Li and $^8$B  only by 2$\%$.

The values $b_{lj}^2$ calculated with the V2 and MN potentials
are presented in Table 1 as well. 
The ratio of $b_{lj}^2$ for the mirror
overlaps differs by 5 to 10$\%$
from the single-particle  estimate ${\cal R}_{s.p.}$ = 
1.01$\pm$0.01  obtained
on the
assumption that   mirror single-particle potential wells are exactly
the same. Therefore, the present MCM calculations suggest that this assumption
is not valid.

It is interesting to note that 
MCM predicts that 
the r.m.s. radius $\la r_{1\frac{1}{2}}^2 \ra ^{1/2}$ 
should be larger than $\la r_{1\frac{3}{2}}^2 \ra ^{1/2}$.
The same result has been obtained earlier in Ref. \cite{Tim} where  the
overlap integrals were found as solutions of the inhomogeneous equation 
with a shell model source term.  
A standard single-particle potential model with  
central  and  spin-orbit potentials
predicts that the  single-particle wave function with $j = 3/2$ 
has smaller radius than the one with $j = 1/2$.
To achieve this inversion of the r.m.s. radii, the phenomenological 
single-particle
spin-orbit potential should be taken with opposite sign.  
Understanding the differences in $j=3/2$ and $j=1/2$ overlaps is important
for the accurate determination of ANC from transfer reactions.


\subsubsection{$^{12}{\rm B} - ^{12}{\rm N}$}

To study the overlap integrals $\la ^{12}$N$|^{11}$C$\ra$ and
$\la ^{12}$B$|^{11}$B$\ra$ 
we use the wave functions of $^{12}$B and $^{12}$N calculated earlier
in Ref. \cite{Des99} in the    multichannel
two-cluster  models $^{11}$B + n and $^{11}$C + p with
excited states $\frac{1}{2}^-$, $\frac{3}{2}^-$,
$\frac{5}{2}^-$ and $\frac{7}{2}^-$ of the $^{11}$B and $^{11}$C cores
 taken into account.

\begingroup
\squeezetable
\begin{table*} 
\caption{Asymptotic normalization coefficients 
squared
$C_{lj}^2$ (in fm$^{-1}$), their sums $C_l^2$, 
spectroscopic factors $S_{lj}$ and $S_l = S_{l,l-\frac{1}{2}} + 
S_{l,l+\frac{1}{2}}$, ratio $b_{lj}^2 = C_{lj}^2/S_{lj}$ (in fm$^{-1}$)
and r.m.s. radii $\la r^2_{lj}\ra^{1/2}$ (in fm)
for mirror  overlap integrals. The calculations have been performed 
with
two NN potentials, V2 and MN.
The ratios ${\cal R}$ of similar quantities of mirror overlaps are  
given in each third line. 
 } 
\begin {center}
\begin{tabular}{  p{0.8 cm}
p{0.8 cm} p{0.8 cm} p{1.6 cm} p{1.6 cm}  p{1.4 cm} p{1.3 cm} p{1.3 cm} 
p{1.3 cm}  p{1.3 cm}  p{1.3 cm}  p{1.6 cm}  p{1.6 cm} }
\hline 
\\
 & & $l$ & $C_{l,l-\frac{1}{2} }^2$ & $C_{l,l+\frac{1}{2} }^2$ & $C_{l}^2$ 
& $S_{l,l- \frac{1}{2} }$ & $S_{l,l+ \frac{1}{2} }$ & $S_l$ &
$b_{l,l- \frac{1}{2} }^2$ & $b_{l,l+ \frac{1}{2} }^2$ & 
$\la r^2_{l,l- \frac{1}{2} }\ra^{1/2}$ & 
$\la r^2_{l,l+ \frac{1}{2} }\ra^{1/2} $ \\
 & & & &  & & & & & & & \\
\hline

\multicolumn{12}{c}{$\la^{8}{\rm B}(2^+)|^{7}{\rm Be}
(\frac{3}{2}^-)\ra - \la^{8}{\rm Li}(2^+)|^{7}{\rm Li}(\frac{3}{2}^-)\ra$} \\

  & $p$ &1 &  0.0886 & 0.6850 & 0.7736 & 0.104 & 0.926 
  & 1.030 & 0.850 & 0.740 & 5.08 & 4.83 \\
V2 & $n$ &1  & 0.0706 & 0.6539 & 0.7244  & 0.086 & 0.955 
& 1.040 & 0.823 & 0.685 & 4.078 & 3.87 \\
 &   ${\cal R}$ & & 1.256 & 1.048 & 1.068  & 1.217 & 0.970 
 & 0.990 & 1.033 & 1.098 & & \\
  \hline
  & $p$ &1 & 0.0811 & 0.5602 & 0.6413 & 0.114  & 0.922 & 1.037 
  & 0.709 & 0.607 & 4.76  & 4.51 \\
MN  & $n$ & 1 & 0.0682 & 0.5193 & 0.5875 & 0.102  & 0.942& 1.044 & 0.668 
& 0.552 & 3.83  & 3.64\\
 & ${\cal R}$ &  & 1.189 & 1.079 & 1.092 & 1.119 & 0.979 & 0.993 & 1.06  
 & 1.10 & & \\
  \hline
  \hline
  \multicolumn{12}{c}{$\la^{12}{\rm N}(1^+)|^{11}{\rm C}
  (\frac{3}{2}^-)\ra - \la^{12}{\rm B}(1^+)|^{11}{\rm B}(\frac{3}{2}^-)\ra$} \\
  & $p$ &1 & 1.76 & 0.595 & 2.35 & 0.672 & 0.261 & 0.933 & 2.61 & 2.28 
  & 4.07 & 3.92 \\
V2  & $n$ & 1 & 1.38 & 0.445 & 1.82 & 0.684 & 0.262 & 0.946 & 2.02 & 1.70 
& 3.59  & 3.47 \\
 &  ${\cal R}$ & & 1.28 & 1.34 & 1.29 & 0.982 & 0.996 & 0.986 & 1.29 
 & 1.34 & & \\
  \hline
  & $p$ &1 & 1.529 & 0.598 & 2.127 & 0.637 & 0.276 & 0.913 & 2.40 & 
  2.17 & 3.97 & 3.86\\
MN   & $n$ & 1 & 1.201 & 0.440 & 1.641 & 0.661 & 0.278 & 0.939 & 
1.82 & 1.58 & 3.50  & 3.41\\
 
 &  ${\cal R}$ &   & 1.27 & 1.36 & 1.30 & 0.964 & 0.993 & 0.972 & 
 1.32 & 1.37 & & \\
  \hline
  \hline
    \multicolumn{12}{c}{$\la^{13}{\rm N}
    (\frac{1}{2}^-)|^{12}{\rm C}(0^+)\ra - 
    \la^{13}{\rm C}(\frac{1}{2}^-)|^{12}{\rm C}(0^+)\ra$  
     two-cluster model} \\
    & $p$ &1 & & 2.66& &  & 0.530 &  & & 5.01 & & 3.63     \\
V2  & $n$ & 1 & & 2.36& &  & 0.531 &  & & 4.45 & & 3.37     \\
    & ${\cal R}$ &  & & 1.13& &  & 0.998 &  & & 1.13 & &          \\

  \hline
    & $p$ &1 & & 2.18 &&  & 0.502 & & & 4.35 & & 3.50    \\
MN  & $n$ & 1 & & 1.92 &&  & 0.498 & & & 3.85 & & 3.26    \\      
    & ${\cal R}$ &  & & 1.14 &&  & 1.008 & & & 1.13 & &  \\
\hline  
\hline
    \multicolumn{12}{c}{$\la^{13}{\rm N}
    (\frac{1}{2}^-)|^{12}{\rm C}(0^+)\ra - 
    \la^{13}{\rm C}(\frac{1}{2}^-)|^{12}{\rm C}(0^+)\ra$  
     four-cluster model} \\
    & $p$ &1 & & 1.54$\pm$0.04& &  & 0.335 &  & & 4.61$\pm$0.10 & & 3.58   \\
V2  & $n$ & 1 & & 1.30$\pm$0.04& &  & 0.330 &  & & 3.95$\pm$0.12 & & 3.32   \\
    & ${\cal R}$ &  & & 1.19$\pm$0.01& &  & 1.01 &  & & 1.17$\pm$0.01 & &  \\

  \hline
    & $p$ &1 & & 1.34$\pm$0.04 &&  & 0.341 & & & 3.93$\pm$0.12 & & 3.44   \\
MN  & $n$ & 1 & & 1.12$\pm$0.06 &&  & 0.336 & & & 3.33$\pm$0.18 & & 3.20  \\ 
    & ${\cal R}$ &  & & 1.19$\pm$0.01 &&  & 1.01 & & & 1.17$\pm$0.01 & &  \\
\hline
\hline
    \multicolumn{12}{c}{$\la^{15}{\rm O}
    (\frac{1}{2}^-)|^{14}{\rm N}(1^+)\ra - 
    \la^{15}{\rm N}(\frac{1}{2}^-)|^{14}{\rm N}(1^+)\ra$  
    } \\
  & $p$ &1 & 64.7 & 0.830 & 65.5 & 1.420 & 0.017 & 1.437 & 45.6 
  & 48.8 & 3.10 & 3.15\\
V2  & $n$ & 1 & 43.9 & 0.568 & 44.5 & 1.456 & 0.017 & 1.473 & 30.2 
& 33.27 & 3.00 & 3.05 \\
 &  ${\cal R}$ & & 1.473 &  1.461 & 1.473 & 0.975 & 1.006 & 0.976 
 & 1.511 & 1.465 &  &\\
  \hline
  & $p$ &1 & 52.7 & 0.051 & 52.7 & 1.465 & 8.6$\times 10^{-4}$ & 1.466 
  & 35.9 & 58.6 & 2.98 & 3.35\\
MN   & $n$ & 1 & 35.6 & 0.036 & 35.6 & 1.489 & 8.2$\times 10^{-4}$ 
& 1.489 & 23.9 & 43.7 & 2.89 & 3.29\\
 
 & ${\cal R}$ &  & 1.479 & 1.417 & 1.481 & 0.984 & 1.049 & 0.985 & 
 1.502 & 1.34 & & \\
     
\hline
\hline  
    \multicolumn{12}{c}{$\la^{15}{\rm O}
    (\frac{3}{2}^+)|^{14}{\rm N}(1^+)\ra - 
    \la^{15}{\rm N}(\frac{3}{2}^+)|^{14}{\rm N}(1^+)\ra$  
    } \\ 
   & $p$ &0 &  &    33.18 & &  & 0.986 &  &  & 33.7 &  & 5.01 \\
V2  & $n$ & 0 &   & 8.74 &   &   & 0.953 &   &   & 9.17 &  & 4.15 \\
  &  ${\cal R}$ & &  &    3.79 & &  & 1.035 &  &  & 3.68 & & \\
  \hline
   & $p$ &0 &  & 29.4 &   &  & 0.995 &  &  & 29.6 &  & 4.80 \\
MN  & $n$ & 0 &  & 7.67 &   &  & 0.966 &  &   &  7.93&   & 3.99 \\
 & ${\cal R}$ &  &  &    3.82 & &  & 1.03 &  &  & 3.73 & & \\
\hline
\hline
    \multicolumn{12}{c}{$\la^{17}{\rm F}
    (\frac{5}{2}^+)|^{16}{\rm O}(0^+)\ra - 
    \la^{17}{\rm O}(\frac{5}{2}^+)|^{16}{\rm O}(0^+)\ra$ }\\
    & $p$ &2 & & 1.09 & & & 1.122 & & & 1.056 & & 3.84\\
V2  & $n$ & 2 & & 1.00 & & & 1.125 & & & 0.889 & & 3.61  \\
    &  ${\cal R}$ & & & 1.19 & & & 0.997 & & & 1.19  & & \\
  \hline
    & $p$ &2 & & 0.951& & & 1.124 & & & 0.846 & & 3.67 \\
MN  & $n$ & 2 & & 0.796 && & 1.126 & & & 0.706 & & 3.47  \\
    & ${\cal R}$ &  & & 1.19 & & & 0.998 & & & 1.20  & & \\
  \hline
  \hline
 
      \multicolumn{12}{c}{$\la^{17}{\rm F}
    (\frac{1}{2}^+)|^{16}{\rm O}(0^+)\ra - 
    \la^{17}{\rm O}(\frac{1}{2}^+)|^{16}{\rm O}(0^+)\ra$ }\\
    & $p$ &0 & & 8000 & & & 1.095 & & & 7277 & &  5.55\\
V2  & $n$ & 0 & & 11.0 & & & 1.110 & & & 9.93 & & 4.40  \\
    &  ${\cal R}$ & & &  727 & & & 0.986 & & & 733 & & \\
  \hline
    & $p$ &0 & & 7110 & & & 1.110 & & & 6444 & & 5.32 \\
MN  & $n$ & 0 & & 9.66 & & & 1.113 & & & 8.68 & & 4.24 \\
    & ${\cal R}$ &  & &  736 & & & 0.997 & & & 742  & & \\

\hline
\hline
          \multicolumn{12}{c}{$\la^{23}{\rm Al}
    (\frac{5}{2}^+)|^{22}{\rm Mg}(0^+)\ra - 
    \la^{23}{\rm Ne}(\frac{5}{2}^+)|^{22}{\rm Ne}(0^+)\ra$ }\\
    & $p$ &2 & &  1.17$\times 10^4$ & & & 0.285 & & & 4.12$\times 10^4$ & 
    & 3.93\\
V2  & $n$ & 2 & &  0.398 & & & 0.299 && & 1.33 & & 3.68 \\
 & ${\cal R}$ & &  & 2.95$\times 10^4$  & &  & 0.953 & & &  3.11$\times 10^4$ 
  & & \\
  \hline
  & $p$ &2 & & 1.02$\times 10^4$ & & & 0.281 & & & 3.61$\times 10^4$ & & 3.83\\
MN  & $n$ & 2 & & 0.343 & &  & 0.294 & & & 1.17 & & 3.60 \\
 &  ${\cal R}$ & &  &  2.96$\times 10^4$ & & & 0.956 & & &  3.09$\times 10^4$ 
 & &\\
  \hline
  \hline
            \multicolumn{12}{c}{$\la^{27}{\rm P}
    (\frac{1}{2}^+)|^{26}{\rm Si}(0^+)\ra - 
    \la^{27}{\rm Mg}(\frac{1}{2}^+)|^{26}{\rm Mg}(0^+)\ra$ }\\
    & $p$ &0 &  & 1648 & & & 0.901 & & & 1830 & & 4.43\\
V2  & $n$ & 0 &  & 36.0  & & & 0.824 && &  45.2 & & 3.93 \\
    &  ${\cal R}$ & &  & 45.8  & &  & 1.09 & & &  40.5 &  & \\
  \hline
    & $p$ &0 & & 1380  & & & 0.873 & & &  1582 & & 4.28\\
MN  & $n$ & 0 & & 31.1  & & & 0.809 & & & 38.5 & & 3.81 \\
    & ${\cal R}$ &  &  &  44.3 & & & 1.08 & & &  41.1 & & \\
  \hline
 \hline
          
\end{tabular}
\end{center}
\label{table1}
\end{table*} 
\endgroup

The ANCs, spectroscopic factors, r.m.s. radii of these overlaps and
 the single-particle ANCs $b_{lj}$  are presented in Table I.
The dependence of these values on the  NN potential choice is   weaker 
 than in the case of $^8$Li-$^8$B.
The ratio  ${\cal R}_{MCM}$  
depends on the NN potential choice 
less than the ANCs themselves,  and the difference
between ${\cal R}_{\frac{3}{2}}$  
and ${\cal R}_{\frac{1}{2}}$ is smaller than for the $^8$B - $^8$Li mirror pair. 
${\cal R}_{MCM}$, which is equal to 1.29 for V2 and 1.30 for MN,
  agrees well with the single-particle estimate ${\cal R}_{s.p.} =
1.30 \pm 0.02$ obtained on the assumption of charge-symmetry of
mirror  single-particle potential wells. However, it is smaller
than the prediction ${\cal R}_0$ = 1.38 $\pm$ 0.02 
of Eq. (\ref{rn}) by $6$\%.
In section IV.A we have shown that for the $p$-shell nucleus $^8$B with the 
proton separation energy  similar to that in $^{12}$N,
  ${\cal R}_{MCM}$ agrees with ${\cal R}_0$ within uncertainties of the
calculation of the latter (see $m$ = 0.56 and $u$ = 1.01
cases in Fig.1). Therefore, the 6$\%$ deviation of ${\cal R}_{MCM}$   
from ${\cal R}_0$, obtained in this section, can be attributed
to the charge symmetry breaking in the effective NN interactions,
which is about 1.9$\%$ for V2 and 5.8$\%$ for MN.

The neutron ANC $C_l^{exp}$ = 1.16 $\pm$ 0.10 fm$^{-1/2}$ 
and the r.m.s. radius $\la r^2_{exp} \ra^{1/2}$ = 3.16 $\pm$ 0.32 fm
for $\la^{12}$B$|^{11}$B$\ra$ 
have been reported in Ref. \cite{Liu} where they have been determined
from the $^{11}$B(d,p)$^{12}$B reaction. 
Our MCM calculations give the larger values,
 $C_l$ = 1.35 fm$^{-1/2}$ for V2 and 
and  $C_l$ = 1.28 fm$^{-1/2}$ for MN, while
the  theoretical  r.m.s. radius ranges from 3.41   to 3.59 fm 
depending on $j$ and NN force.

The proton  ANCs for $^{12}$N have been determined from the peripheral
transfer reaction $^{14}$N($^{11}$C,$^{12}$N)$^{13}$C in Ref. \cite{XT} 
resulting in $C_{l\frac{1}{2}}^2$ = 1.4 $\pm$ 0.2 fm$^{-1/2}$,
$C_{l\frac{3}{2}}^2$ = 0.33 $\pm$ 0.05 fm$^{-1/2}$ and
$C_{l}^2=C_{l\frac{1}{2}}^2+C_{l\frac{3}{2}}^2$ =1.73 $\pm$ 0.25 fm$^{-1/2}$.
Our MCM calculations overestimate the
experimental  $C_l^2$ value by 35$\%$ for V2 and 23 $\%$
for  MN. The theoretical 
ratio $C_{l\frac{3}{2}}^2/C_{l\frac{1}{2}}^2$ 
0.34 for V2 and 0.39 for MN, is also larger than the experimental value of 
0.24 $\pm$ 0.07.
However, the ANCs in mirror nuclei are overestimated in the same proportion,
so that the theoretical ratio ${\cal R}_{MCM}$ of 1.29 and 1.30 agrees
well with the experimental value ${\cal R}_{exp}$ = 1.28 $\pm$ 0.29.

The spectroscopic factors  in  $^{12}$N and $^{12}$B
change by no more than 6$\%$ with different choices of the NN potential.
The  mirror spectroscopic factors $S_{1\frac{1}{2}}$, 
differ 
by 2.8$\%$ and 3.6$\%$ for V2 and MN respectively while $S_{1\frac{1}{2}}$
are practically the same for both of them.


\subsubsection{$^{13}{\rm C} - ^{13}{\rm N}$}

To describe the mirror pair $^{13}$N - $^{13}$C, we used two different models:
the multichannel two-cluster model $^{12}$C + n(p) from
Ref. \cite{TBD97} and the multichannel four-cluster model
$\alpha+\alpha+\alpha$ + n(p), that has been developed in Ref. \cite{Duf97}.
Numerical precision of ANCs squared obtained in the latter model is
 about 2-3$\%$.
The results of calculations are presented in Table I.

The ANCs obtained in  two-cluster and four-cluster models 
differ by 60 to 80$\%$ and
the spectroscopic factors differ
by about 50 to 60$\%$ depending on the NN potential
used in calculations. Such a large difference arises because
 the 
$\alpha+\alpha+\alpha$ model for the nucleus 
$^{12}$C contains only one type of
permutational symmetry determined by the Young diagram  $[f]$ = [444]. 
 As explained in Ref. \cite{TBD97}, the main contribution
to the spectroscopic factor, vertex constant, and therefore, to the ANC 
of the overlap integral $\la ^{13}$C$|^{12}$C$\ra$
comes from 
the overlap between the  [4441]$^{22}$P state in $^{13}$C
and the [4431]$^{13}$P  state in   $^{12}$C. 
The  [4431]$^{13}$P  configuration  is absent in the 
$\alpha+\alpha+\alpha$ model
but  is present in the one-center shell model wave function of $^{12}$C
used in the  two-cluster model.
For this reason, the two-cluster model gives larger
ANCs, and spectroscopic factors for $^{13}$C
and $^{13}$N, than the  four-cluster model.

Several experimental values for the neutron ANC of $^{13}$C are available
\cite{G77,K88,P71,G82,gala,Liu}. Apart from the latest value from
Ref. \cite{Liu}, obtained from a  non-peripheral (d,p) reaction,
they agree with each other leading to  an average  value
$C^2_l$ = 2.36 $\pm$ 0.12 fm$^{-1}$. Our two-cluster calculations with V2
agree with this value while the same calculations performed with  MN  
underestimate it. However, such calculations are very sensitive 
to the spin-orbit force, as it regulates the probability of the [4431]$^{13}$P
configuration in $^{12}$C \cite{TBD97}. 
As for the four-body model, it underestimates
the experimental values $C_l^2$ squared by a factor of two.

The ratio ${\cal R}_{MCM}$, calculated in the four-cluster model agrees
well both with the analytical value ${\cal R}_0$ = 1.198 $\pm$ 0.004 and
the single-particle value ${\cal R}_{s.p.}$ = 1.168 $\pm$ 0.020. 
However, the two-cluster
model gives smaller values of ${\cal R}_{MCM}$, 1.13 and 1.14 for
the V2 and MN potentials respectively. As we have seen in
Sec. IV.A, the two-cluster model ${\cal R}_{MCM}$
agrees  both with ${\cal R}_0$ and ${\cal R}_{s.p.}$ if
the charge symmetry of the NN interactions is present. In this section,
to reproduce the mirror separation energies $\epsilon_p$ and $\epsilon_n$
within the two-cluster model, the Majorana
parameters $m$ of V2
in mirror nuclei   $^{13}$C and $^{13}$N have
to be different by 1.4$\%$ and the parameters $u$ of MN
must differ by 1.9$\%$, which corresponds to   $\sim$6$\%$ difference
in the odd NN potentials. With actual parameters $m$ and $u$,
used in the two-cluster calculations, the   singlet- and triplet-odd parts 
of the NN potentials
are large. As a result, the deviation of ${\cal R}_{MCM}$ from what
would be expected in the case of   charge symmetry, 
is comparable to the degree in which  charge symmetry 
is broken. 
The situation is different for the four-cluster model  where
the required charge symmetry breaking 
in these components is also smaller
and the actual choice of parameters $m$ and $u$ gives to  weaker 
odd NN potentials.

The spectroscopic factors obtained  are sensitive both to the model 
and the NN potential choice, however the difference in
mirror spectroscopic factors does not exceed 2$\%$.  The same model that
reproduces the experimental ANC value in $^{13}$C  gives
the spectroscopic factor $S$ = 0.53 which is lower than the shell
model value of 0.68 of Ref. \cite{CoKu}.


\subsubsection{$^{15}{\rm O} -  ^{15}{\rm N}$}

We describe   $^{15}$O and $^{15}$N in the multichannel
two-cluster $^{14}$N + p(n) model with the core $^{14}$N being either in
the ground state or in one of the first excited states 1$^+$,   2$^+$  
or 3$^+$. The internal structure of the $^{14}$N core is represented
by the 0$p$ oscillator shell model with the oscillator radius of 1.6 fm. 
We  consider
only two states in $^{15}$O and $^{15}$N, the ground state and the first
$3/2^+$ state, since they are the most important for
understanding   $^{15}$O production in the CNO cycle.

For the ground states of $^{15}$O and $^{15}$N,  
$|C_{1\frac{3}{2}}|^2$ are about two
orders of magnitude  smaller than $|C_{1\frac{1}{2}}|^2$ for both NN potentials
used in the calculations (see Table I), while the
experimentally determined $|C_{1\frac{3}{2}}|^2$ is only one tenth of
$|C_{1\frac{1}{2}}|^2$ \cite{Muk03}. The   $|C_l|^2$  = 65.5 fm$^{-1}$
value  calculated with V2 agrees with experimental value of
63 $\pm$ 14 fm$^{-1}$ from Ref. \cite{Ber02}, while $|C_l|^2$ = 52.7 fm$^{-1}$ 
calculated with MN   agrees with another available experimental value 
of 54 $\pm$ 5.9 fm$^{-1}$ \cite{Muk03}.  The ratio ${\cal R}_{MCM}$ = 1.48, 
which is  almost the same 
for both NN potentials, agrees well
with the analytical value  ${\cal R}_0$ =
1.48 obtained from Eq. (\ref{rn}) and with single-particle value
${\cal R}_{s.p.}$ = 1.51 $\pm$ 0.03. 

The difference between  mirror spectroscopic factors does not exceed
2.5$\%$ for $j$ = 1/2, but it is slightly larger for $j$ = 3/2 
and the MN potential. This difference is most likely due to the 
$\sim 3\%$ difference in the NN potential parameters in mirror states
required for simultaneous reproduction
of proton and neutron separation energies in $^{15}$O and $^{15}$N.

The ANCs for the first excited $3/2^+$ state is less sensitive to the
NN potential choice than those for the ground state. 
An experimentally determined value $C_l^2$ = 21 $\pm$ 5 fm$^{-1}$ for
the $\la ^{15}$O$(\frac{3}{2}^+_1)|^{14}$N$\ra$ has been
reported in Ref. \cite{Ber02}. The experimental data from this work have been
recently reanalysed  in Ref. \cite{Muk03},
increasing this value to  $C_l^2$ = 27.6 $\pm$ 6.8 fm$^{-1}$. The results of our
calculations, 33.2 and 29.4 fm$^{-1}$, are close to this reconsidered value.

The   ${\cal R}_{MCM}$ values for  3/2$^+_1$, calculated with V2 and MN,
  differ only by 1$\%$ and
this value, ${\cal R}_{MCM}$ = 3.8, is smaller than the analytical
estimate  ${\cal R}_0$ = 4.23 $\pm$ 0.15 from Eq. (\ref{rn})
but larger than the single-particle value ${\cal R}_{s.p.}$ = 3.62 $\pm$ 0.03.
This difference must originate purely to the charge symmetry breaking
due to the Coulomb interaction since the   parameters $m$ and $u$ of
nuclear NN potentials differ  less than by half of a per cent
in the mirror 3/2$^+$ states. The Coulomb effects should be also responsible for
 3$\%$ difference in mirror spectroscopic factors and
for deviation of  $(b_p/b_n)^2$   from the single-particle value
${\cal R}_{s.p.}$ = 3.62 $\pm$ 0.03.


\subsubsection{$^{17}{\rm F} -  ^{17}{\rm O}$}

To describe   $^{17}$F and $^{17}$O, we use 
 single-channel two-cluster  models $^{16}$O + n and $^{16}$O + p  
from Refs.
\cite{BT92,BDH98}.  To reproduce simultaneously the proton and neutron 
separation energies in  $^{17}$F and $^{17}$O,  less than 1$\%$ difference in
the NN potential parameters  in mirror states is required.

The ANCs calculated with V2 are on average 13-14$\%$ larger than those
obtained with the MN potential (see Table I). However, 
the ratio ${\cal R}_{MCM}$ of mirror ANCs does not change 
with NN potential choice in  the ground states and
differs  only by 1$\%$ in the first excited states.
The spectroscopic factors are practically insensitive to the NN potential and
differ in mirror states by approximately 1$\%$.

In the $^{17}$F and $^{17}$O ground states,  the   ${\cal R}_{MCM}$ = 1.19 value
agrees with the single-particle estimate ${\cal R}_{s.p.}$ = 1.21 $\pm$ 0.03 
based on charge symmetry of
mirror potential wells and is slightly smaller than prediction
${\cal R}_0$ = 1.21 from the analytical formula (\ref{rn}). However,
for the first excited state 1/2$^+$,
${\cal R}_{MCM}$ $\approx$ 730  is noticably larger than  the
single-particle value of 
702$\pm$4 and significantly smaller than the analytical
value ${\cal R}_0$ 
= 837 $\pm$ 42. In Sec.IV.A we have shown that, in the presence of charge
symmetry of the NN interactions,
the ${\cal R}_{MCM}$ value, calculated  
for very small proton separation energies,
 is approximately the average between ${\cal R}_0$ 
and ${\cal R}_{s.p.}$. The ${\cal R}_{MCM}$ value of the present section
is about 6$\%$ smaller than (${\cal R}_0$+${\cal R}_{s.p.}$)/2 which
should be due to the charge symmetry breaking required to reproduce
mirror separation energies $\epsilon_p$ and $\epsilon_n$ in the $1/2^+$ state.

The ratio $b^2_p/b^2_n$  of mirror  single-particle ANCs squared (733 for V2
and 742 for MN) for the
first excited state 1/2$^+$ is larger  than ${\cal R}_{s.p.}$.
This means that
in the effective local two-body potential model, 
the nuclear potential fields for 1$s_{\frac{1}{2}}$ protons and neutrons
are slightly different.  This contrasts with 
the situation for 0$d_{\frac{5}{2}}$ proton and neutron in ground states of
$^{17}$F and $^{17}$O, where they
can be considered as being placed in  the same nuclear potential well.

The results of the calculations described above 
have been obtained  with an oscillator radius 
of 1.76 fm which reproduces the r.m.s. radius of $^{16}$O. We have repeated
the same calculations  with much smaller value of the oscillator 
radius, $r_0$ = 1.5 fm, in order
to check how   ${\cal R}_{MCM}$ depends on the wave function of the
core $^{16}$O.
With   smaller $r_0$,   $^{16}$O has a 38$\%$ smaller r.m.s. radius, 
the expectation energy of the $^{16}$O
core  is lowered  by 20 MeV   and   $C_l^2$ drops by about 40$\%$. 
However, the
${\cal R}_{MCM}$ changes only by 2$\%$   and   5$\%$  
for the 5/2$^+$ and 1/2$^+$ states respectively. 
This is consistent with the idea behind 
the formula (\ref{rn}) that the ratio of mirror ANCs   depends only on
the core charge and on the separation energies of mirror proton and neutron.

The experimental value $C_l^2$ = 0.667 $\pm$ 0.042 fm$^{-1}$ 
for $^{17}$O$_{g.s.}$ has been 
determined  in Ref. \cite{bz}. As   already reported in Ref. 
\cite{BT92}, the MCM calculations with V2 and MN overestimate this value.
For the mirror nucleus $^{17}$F, the proton ANC has been experimentally 
determined 
in Refs. \cite{fort,vern,art1,art2,gagl} (the ANCs from the data measured
in \cite{fort,vern} are
given in \cite{art2}). The $C_l^2$ values from the first
four works,  0.772 $\pm$ 0.19,  0.911 $\pm$ 0.082, 0.811 $\pm$ 0.082 and
0.838 $\pm$ 0.05 fm$^{-1}$,  agree with each other within the error bars
giving the average value of 0.836 $\pm$ 0.050 fm$^{-1}$. 
However, the $C_l^2$ = 1.08 $\pm$ 0.10 fm$^{-1}$ from
Ref. \cite{gagl} is about 30$\%$ larger.
The theoretical value  ${\cal R}_{MCM}$ = 1.19 agrees well with the 
averaged experimental 
value ${\cal R}^{exp} = (C^{exp}_p/C^{exp}_n)^2$ = 1.25 $\pm$ 0.15
if the ANC from Ref. \cite{gagl} is disregarded.


\begin{figure}[t]
\centerline{\psfig{figure=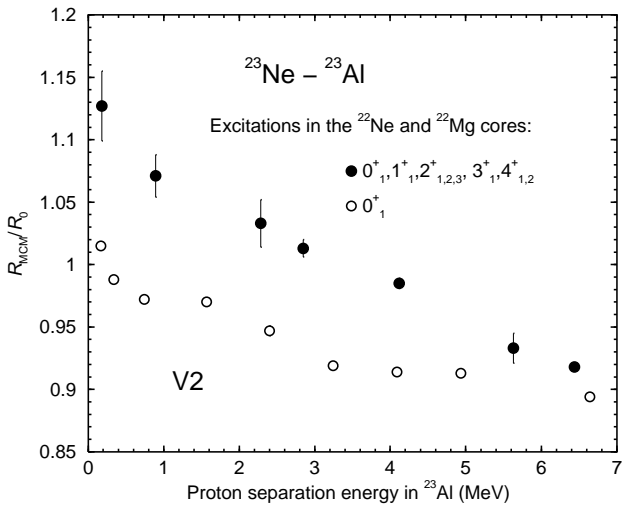,width=0.46\textwidth}}
\caption{Ratio ${\cal R}_{MCM}/{\cal R}_0$ for the $^{23}$Al - $^{23}$Ne
mirror pair 
as a function of the proton separation energy in
$^{23}$Al calculated   without (filled circles) and with (open circles)
excitations in the $^{22}$Mg and $^{22}$Ne cores. The error bars are due 
to uncertainties in calculating ${\cal R}_0$, as explained in Sect. IV A.
The calculations have been performed with different Majorana
parameters $m$ of the Volkov potential V2. Experimental proton 
separation energy is 0.123 MeV. 
}
\end{figure}

\subsubsection{$^{23}{\rm Al} - ^{23}{\rm Ne}$}

\begin{table*}[t]
\caption{  ${\cal R}_{MCM}$ and  ${\cal R}_0$ for the $^{23}$Ne - $^{23}$Al
and $^{27}$P - $^{27}$Mg
 mirror pairs
calculated with different excitations in the   $^{22}$Ne - $^{22}$Mg 
and  $^{26}$Si - $^{26}$Mg  cores.
The calculations have been performed with Volkov potential V2 assuming the same
interactions in mirror nuclei and with Majorana parameters $m$ chosen
to fit  the experimental proton separation energies in $^{23}$Al or $^{27}$P.} 
\begin {center}
\begin{tabular}{ p{7 cm} p{3.0 cm} p{3 cm} p{3 cm}  }
\hline 

Core excitations
& ${\cal R}_{MCM}$ &
${\cal R}_0$ & ${\cal R}_{MCM}/{\cal R}_{0}$   \\
\hline
 &    \multicolumn{3}{c}{$\la^{23}{\rm Al}
    (\frac{5}{2}^+)|^{22}{\rm Mg}(0^+)\ra - 
    \la^{23}{\rm Ne}(\frac{5}{2}^+)|^{22}{\rm Ne}(0^+)\ra$ }\\ 
0$^+_1$ & 1.86$\times 10^4$ & 1.82$\times 10^4$ & 1.02 \\ 
0$^+_1$,2$^+_1$ & 2.24$\times 10^4$ & 2.15$\times 10^4$ & 1.05 \\ 
0$^+_1$,2$^+_1$,4$^+_1$ & 2.90$\times 10^4$ & 2.67$\times 10^4$ & 1.09 \\
0$^+_1$,1$^+_1$,2$^+_1$,4$^+_1$ & 2.93$\times 10^4$ & 2.67$\times 10^4$ 
& 1.10 \\
0$^+_1$,1$^+_1$,2$^+_1$,3$^+_1$,4$^+_1$ & 
       2.96$\times 10^4$ & 2.65$\times 10^4$ & 1.12 \\
0$^+_1$,1$^+_1$,2$^+_{1,2}$,3$^+_1$,4$^+_1$ & 
       3.17$\times 10^4$ & 2.84$\times 10^4$ & 1.12 \\
0$^+_1$,1$^+_1$,2$^+_{1,2,3}$,3$^+_1$,4$^+_1$ & 
       3.21$\times 10^4$ & 2.87$\times 10^4$ & 1.12 \\
0$^+_1$,1$^+_1$,2$^+_{1,2,3}$,3$^+_1$,4$^+_{1,2}$ & 
       3.86$\times 10^4$ & 3.37$\times 10^4$ & 1.15 \\
       \hline \hline
   &                \multicolumn{3}{c}{$\la^{27}{\rm P}
    (\frac{1}{2}^+)|^{26}{\rm Si}(0^+)\ra - 
    \la^{27}{\rm Mg}(\frac{1}{2}^+)|^{26}{\rm Mg}(0^+)\ra$ }\\
0$^+_1$ & 44.04 & 46.3 & 0.95 \\ 
0$^+_1$,2$^+_1$ & 46.96 & 45.7 & 1.03 \\ 
0$^+_1$,2$^+_1$,4$^+_1$ & 47.08 & 45.8 & 1.03 \\
\hline    
\end{tabular}
\end{center}
\label{table2}
\end{table*}

To check if the relation between mirror ANCs is still valid with increasing
mass and charge of a mirror pair, we have calculated the overlap integrals
$\la^{23}$Ne$(\frac{5}{2}^+)|^{22}$Ne$(0^+)\ra$ and 
$\la^{23}$Al$(\frac{5}{2}^+)|^{22}$Mg$(0^+)\ra$. 
The latter is relevant to the proton capture reaction 
$^{22}$Mg(p,$\gamma)^{23}$Al in novae \cite{Cag01}.

We describe  $^{23}$Al and $^{23}$Ne in the multichannel two-cluster
models $^{22}$Mg + p and $^{22}$Ne + n respectively
where the cores $^{22}$Mg and
$^{22}$Ne are in the ground state 0$^+$ and in the excited
$1^+_1$, $2^+_{1,2,3}$, 3$^+_1$ and 4$^+_{1,2}$ states.
The internal structure of these states is
represented by closed $0s$ and $0p$ shells and
linear combinations of all possible Slater determinants of the 
$0d_{\frac{5}{2}}$ shell 
with the oscillator radius   chosen to be 1.7 fm.
The results of these  calculations are presented in Table I.

The calculated ratio ${\cal R}_{MCM}$  $\approx $
2.95$\times 10^4$ is about 12$\%$ higher than both the analytical value 
${\cal R}_0$ = (2.64 $\pm$ 0.03)$\times 10^4$ and the
single-particle value 
${\cal R}_{s.p.}$ = (2.67 $\pm$ 0.03)$\times 10^4$. 
It is unlikely that  such a deviation could come from the 1.5$\%$
difference in the NN potential parameters needed to reproduce both proton
and neutron separation energies  
in $^{23}$Al and $^{23}$Ne. To exclude this reason, we
have computed ${\cal R}_{MCM}$ 
using exactly the same NN interactions in these mirror nuclei. 
As the result, the divergence
between  ${\cal R}_{MCM}$ and  ${\cal R}_0$   has increased
and reached 15$\%$. 
The agreement between 
${\cal R}_{MCM}$ and ${\cal R}_0$  
has been restored 
after  we have dropped all  channels but one, namely,
 $^{22}$Mg$(0^+_1)$ + p and
$^{22}$Ne$(0^+_1)$ + n, in the wave functions of
$^{23}$Al and $^{23}$Ne.
By adding and eliminating different
configurations (see Table II), we have found out that the main reason 
for the difference between  ${\cal R}_{MCM}$ 
and ${\cal R}_0$ 
is  the coupling to the
2$^+_1$ and 4$^+_1$ members of the 0$^+$ ground state rotational band 
and to the second excited state 4$^+_2$ in the $^{22}$Ne
and $^{22}$Mg cores. The  spectroscopic factors calculated in the MCM for
these core excitations, 0.62, 0.75 and 0.95 respectively, are 
significantly larger
than the spectroscopic factors of $\sim$0.29 for the cores  $^{22}$Mg$(0^+_1)$
and $^{22}$Ne$(0^+_1)$ in their ground states.
These spectroscopic factors for mirror overlaps 
$\la^{23}$Ne$(\frac{5}{2}^+)|^{22}$Ne$(0^+)\ra$ and 
$\la^{23}$Al$(\frac{5}{2}^+)|^{22}$Mg$(0^+)\ra$ differ by about 4.5$\%$
and they are reasonably close to the value of 0.34 predicted
by the shell model calculations in Ref. \cite{Cag01}.

Growing  disagreement between ${\cal R}_{MCM}$ and ${\cal R}_0$
with including more  core excitations can be explained by increasing role of
quadrupole term of $\Delta V_{Coul}$ in deformed nuclei.
This term decreases slowly at large $r$ as $r^{-3}$,  giving rise to
contributions to Eq. (\ref{vff}) from beyond the nuclear range $R_N$,
which  were ignored in deriving formula (\ref{rn}) for
${\cal R}_0$.
For very small proton separation energies
the  contribution from nuclear interior to the proton ANC may be
even more reduced  with increasing orbital 
momentum $l$ because of  the $(\kappa r)^l$ behaviour at $r \rightarrow 0$. 
If this is true, then artificial increase of proton separation 
energy in $^{23}$Al should lead to smaller difference between ${\cal R}_{MCM}$
and ${\cal R}_0$. To check this, we have performed the MCM calculations for V2 
with smaller values of $m$. Fig.2 shows that ${\cal R}_{MCM}/{\cal R}_0$ 
indeed decreases with increasing separation energy $\epsilon_p$. The decrease
with $\epsilon_p$, but to a lesser extent, is also present if all
the core excitations are removed (open circles at Fig.2).


\subsubsection{$^{27}{\rm P} - ^{27}{\rm Mg}$}

In this section we study another $sd$-shell mirror pair
$^{27}$P - $^{27}$Mg and the overlap integrals
$\la^{27}$Mg$|^{26}$Mg$\ra$ and 
 $\la ^{27}$P$|^{26}$Si$\ra$.
The latter is relevant to the proton capture reaction 
$^{26}$Si(p,$\gamma)^{27}$P in the $rp$-process in the hot stellar 
hydrogen burning \cite{Her95}.

We describe  $^{27}$P and $^{27}$Mg in the two-cluster
models $^{26}$Si + p and $^{26}$Mg + n respectively
in which the cores $^{26}$Si and
$^{26}$Mg can be in ground state 0$^+_1$ and in first 2$^+_1$ and 4$^+_1$
excited states. The internal structure of these states is
represented by the Slater determinants composed of $0s$, $0p$ and
$0d_{\frac{5}{2}}$ single-particle oscillator wave functions with
the oscillator radius of 1.7 fm.

First,
we have studied the dependence of  
the ratio ${\cal R}_{MCM}/{\cal R}_0$ on core excitations 
using the  assumption of charge-symmetry of the NN interaction. The results, 
presented in Table II, show that coupling
to the configuration  with the core in the 2$^+_1$ state   
increases this ratio by 8$\%$. This configuration  has a
spectroscopic factor of 0.25 which is 3.5 times smaller than that
for the ground state.
These results have been obtained 
for the V2 potential, in which 
the parameter $m$ has been fitted to reproduce the experimental proton
separation energy in $^{27}$P. 

With NN interaction different in mirror nuclei, the difference between
${\cal R}_{MCM}$ and ${\cal R}_0$ is 2.5$\%$. The average value 
${\cal R}_{MCM}$ = 45.0 $\pm$ 0.8 is larger than the single-particle
estimate ${\cal R}_{s.p.}$ = 40.3 $\pm$ 1.1, but ${\cal R}_b=b_p^2/b_n^2 =
40.8\pm 0.3$
agrees with ${\cal R}_{s.p.}$. This means that potential wells
for mirror valence neutron and proton can be considered to be the same.
Therefore, the deviation of ${\cal R}_0$ from ${\cal R}_{s.p.}$ is
due to the difference in mirror spectroscopic factors. This difference, 9$\%$
for V2 and 8$\%$ for MN, is  unexpectedly large.

The average value of the spectroscopic factor in $^{27}$P and $^{27}$Mg, 
which is $\sim$ 0.85, is about twice the 
value predicted by the shell model calculations in \cite{Her95}.
Such a disagreement is most likely caused by neglect of $1s_{\frac{1}{2}}$ and
$0d_{\frac{3}{2}}$ orbitals in the core wave functions.

\section{Summary and conclusions}

\begin{table*}[t]
\caption{Number of nodes $n$, orbital momentum $l$,
proton ($\epsilon_p$) and neutron ($\epsilon_n$) separation energies (in MeV),
single-particle estimate ${\cal R}_{s.p.}$, 
microscopic calculations
${\cal R}_{MCM}$, analytical estimate ${\cal R}_0$
  microscopic
calculations for ${\cal R}_b = C^2_pS_n/(C^2_nS_p)$
for the mirror pairs from   the first column.
Average value between  calculations with V2 and MN potentials is
presented for ${\cal R}_{MCM}$ and ${\cal R}_b$. 
Charge symmetry breaking of NN interactions
is assumed in the MCM calculations.  } 
\begin {center}
\begin{tabular}{ p{1.7 cm} p{0.6 cm} p{0.6 cm}  p{1.0 cm}  
p{0.8 cm}  p{2.6 cm} p{2.2 cm} 
p{2.2 cm}  p{2.2 cm}  p{2.2 cm} }
\hline 
\\ 
Mirror pair  &  $J^{\pi}$ & $\,nl\,$ & $\epsilon_p$ & $\epsilon_n$ 
& ${\cal R}_{s.p.}$  &
${\cal R}_{MCM}$ 
 &  
 ${\cal R}_0$ & \multicolumn{2}{c}{${\cal R}_b$}
 \\
 & & & & &  & & & $j = l-1/2$ & $ j = l+1/2$ \\
\hline

$^{8}{\rm B}  - ^{8}{\rm Li}$ & $2^+$ &

0p & 0.137 &  2.03  &  1.01 $\pm$ 0.01 & 1.075 $\pm$ 0.013 &  1.13 $\pm$ 0.01 &
1.046 $\pm$ 0.014 &  1.099 $\pm$ 0.001 \\

$^{12}{\rm N}  - ^{12}{\rm B}$ & $1^+$ &

0p & 0.601 &  3.37  &  1.30 $\pm$ 0.02 & 1.295 $\pm$ 0.05 &  1.38 $\pm$ 0.02 &
1.305 $\pm$ 0.015 &  1.355 $\pm$ 0.015 \\

$^{13}{\rm N}- ^{13}{\rm C}^{2c}$ & $\frac{1}{2}^-$ &
0p & 1.944 &  4.95 &  1.168 $\pm$ 0.02 & 1.135 $\pm$ 0.005 &  
1.198 $\pm$ 0.004 &
  &  1.13 \\

$^{13}{\rm N} - ^{13}{\rm C}^{4c}$ & $\frac{1}{2}^-$ &
0p & 1.944 &  4.95 &  1.168 $\pm$ 0.02 & 1.19 $\pm$ 0.01 &  1.198 $\pm$ 0.004&
  &  1.17 $\pm$ 0.01 \\

$^{15}{\rm O} - ^{15}{\rm N}$ & $\frac{1}{2}^-$ &
0p & 7.297 &  10.8  &  1.51 $\pm$ 0.03 & 1.477 $\pm$ 0.004 &  1.48 &
1.506 $\pm$ 0.005 &  1.40 $\pm$ 0.06  \\

$^{15}{\rm O}  - ^{15}{\rm N}$ & $\frac{3}{2}^+$ &
1s & 0.507 &  3.53  &  3.62 $\pm$ 0.02 & 3.805 $\pm$ 0.015 &  4.23 $\pm$ 0.15 &
  &  3.705 $\pm$ 0.025 \\

$^{17}{\rm F} - ^{17}{\rm O }$ & $ \frac{5}{2}^+$ &
0d & 0.601 &  4.14  &  1.21 $\pm$ 0.03 & 1.19 &  1.21 &  &  
1.195 $\pm$ 0.005 \\

$^{17}{\rm F}  - ^{17}{\rm O}$ & $\frac{1}{2}^+$ &
1s & 0.106 &  3.27  &  702 $\pm$ 4 &   731 $\pm$ 5 &   837 $\pm$ 42 &
 & 738 $\pm$ 5 \\

$^{23}{\rm Al} - ^{23}{\rm Ne}$ & $\frac{5}{2}^+$ &
0d & 0.123 &  4.42 &  (2.67$\pm$0.03)$\times 10^4$ & 2.95$\times 10^4$ &  
(2.63$\pm$0.03)$\times 10^4$ &  &  
(3.10$\pm$0.01)$\times 10^4$ \\

$^{27}{\rm P} - ^{27}{\rm Mg}$ & $\frac{1}{2}^+$ &
1s & 0.859 &  6.44  &  40.3 $\pm$ 1.1 &  45.0 $\pm$ 0.8 &  
44.0 $\pm$ 0.7 & & 40.8 $\pm$ 0.3 \\
\hline
\multicolumn{7}{l}{$^{2c}$ - two-cluster model} \\
\multicolumn{7}{l}{$^{4c}$ - four-cluster model }\\

\end{tabular}
\end{center}
\label{table3}
\end{table*}

According to the  simple analytical formula (\ref{rn}) derived 
in Ref. \cite{Tim03},
the   ANCs  for mirror virtual nucleon decays are related
because  of charge symmetry of the NN interaction.
This relation is determined 
only by the separation energies of mirror proton and neutron,
the charge of the residual nucleus and the range of its strong interaction  
with the last nucleon. The ratio of mirror ANCs  
is not sensitive to the NN potential 
and  details of internal nuclear structure. This ratio should be
the same in channels with different spin, or for the same
transferred angular momentum $j$.

The MCM calculations of the present paper confirm this general trend.
For the mirror pairs considered here, 
the ratio ${\cal R}_{MCM}$ changes by four orders of magnitude as
 predicted by Eq. (\ref{rn}). Moreover,
when charge symmetry of NN interactions
is  assumed in MCM,  ${\cal R}_{MCM}$ and
${\cal R}_0$ for nodeless overlaps are in good agreement 
even for small separation proton 
energies.  This agreement occurs for
both the NN interactions used in calculations.
For the overlap $\la ^{17}$F$(\frac{1}{2}^+)|^{16}$O$\ra$ with a node,  
a judgement  about the
agreement between ${\cal R}_{MCM}$ and ${\cal R}_0$ is
more difficult to make due to uncertainties in
the choice of $R_N$ to calculate ${\cal R}_0$. Nevertheless,
for very small proton separation energies ${\cal R}_{MCM}$ are more
closer to ${\cal R}_0$ rather than to ${\cal R}_{s.p.}$. 

The most noticeable disagreement between ${\cal R}_{MCM}$ and
${\cal R}_0$ can be seen for small components of overlap integrals, for example,
for $j=1/2$ component in  $\la ^{8}$B$|^7$Be$\ra$. Even in this case,
the disagreement is on the level of $8\%$ if charge symmetry of NN interactions
is valid. Stronger disagreement can occur
for nuclei with deformed cores. For the $^{23}$Al - $^{23}$Ne mirror pair,
strong coupling to the excited states in the $^{22}$Mg and $^{22}$Ne cores  
increase this disagreement up to 15$\%$. 

The charge symmetry breaking of the NN interactions, required
to reproduce simultaneously the experimental proton and neutron 
separation energies, reduces ${\cal R}_{MCM}$  with respect to ${\cal R}_0$.
  This is especially noticeable for
two-cluster calculations of the $^{13}$N - $^{13}$C mirror pair where
this effect reaches  6$\%$ (see Table III). 
These two-cluster calculations require too large
odd NN interactions with strong  breaking of mirror symmetry. 
Four-cluster calculations, which do not require strong breaking
of mirror symmetry, give much better agreement between 
${\cal R}_{MCM}$ and ${\cal R}_0$.  Good agreement between 
${\cal R}_{MCM}$ and ${\cal R}_0$
 also occurs for another 0p overlap $\la ^{15}$O$(\frac{1}{2}^-)|^{14}$N$\ra$.
However with decreasing proton separation energy, for example 
for $^8$B - $^8$Li
and $^{12}$N - $^{12}$N mirror pairs, this agreement
deteriorates and the deviations reach  6$\%$.

For other nodeless overlaps considered here, the agreement between
${\cal R}_{MCM}$ and ${\cal R}_0$ depends on the deformation of the
residual nucleus. In the absence of strong core excitations  (the
$^{17}$F($\frac{5}{2}^+$) - $^{17}$O($\frac{5}{2}^+$) case) 
the agreement between ${\cal R}_{MCM}$  and ${\cal R}_0$  is good, 
however strong coupling to excited states of the core may noticably
increase ${\cal R}_{MCM}$, for example in $^{23}$Al($\frac{5}{2}^+$) -
$^{23}$Ne($\frac{5}{2}^+$). For overlaps with one node and a
loosely bound proton
the situation  is opposite. ${\cal R}_{MCM}$ and ${\cal R}_0$  are in good
agreement if  core excitations are present
($^{27}$P($\frac{5}{2}^+$) - $^{27}$Mg($\frac{5}{2}^+$)),
otherwise ${\cal R}_{MCM}$  is smaller
than ${\cal R}_0$ (as in $^{15}$O($\frac{3}{2}^+$) and 
$^{17}$F($\frac{1}{2}^+$)). 

Our investigation of mirror symmetry of spectroscopic factors has shown that
the spectroscopic factors for small components of one-nucleon overlaps can
differ up to 20$\%$. For large components of overlaps the mirror spectroscopic
factors are almost the same: the spectroscopic factors $S_l = 
S_{l,l-\frac{1}{2}} + S_{l,l+\frac{1}{2}}$ for 0p-shell mirror overlaps
may differ up to 3$\%$.
For single-particle mirror nuclei $^{17}$F and $^{17}$O, the spectroscopic
factors are the same, while for nuclei in the middle of the sd-shell,
mirror spectroscopic factors may differ by up to 9$\%$.

The microscopic calculations of single-particle ANCs 
$b_{lj}=C_{lj}S_{lj}^{-1/2}$ and their ratio squared ${\cal R}_b$ 
for mirror overlaps are presented in Table III where they are compared
to the single-particle estimates based on assumption of charge symmetry
of mirror potential wells. This comparison shows that the concept of
mirror symmetry of potential wells is valid only for $j=1/2$ component
in the  $^{12}$N - $^{12}$B and $^{15}$O$(\frac{1}{2}^-)$ -
$^{15}$N$(\frac{1}{2}^-)$ mirror pairs,   in the 0d nuclei 
$^{17}$F($\frac{5}{2}^+$) and $^{17}$O($\frac{5}{2}^+$)
and for four-cluster calculations of $^{13}$N - $^{13}$C. 
For all other overlap integrals
this assumption is not valid. It is interesting that for first excited
1s-states in $^{17}$F and $^{17}$O, which are supposed to be 
good single-particle nuclei,  ${\cal R}_b$ significantly
differs from ${\cal R}_{s.p.}$. This means that stronger penetration
of the valence 1s neutron inside the $^{16}$O core perturbs the mean field
in greater extent than the  mirror proton leading to mirror symmetry breaking
in single-particle potential wells.

The assumption that in mirror nuclei both mirror potential wells and 
mirror spectroscopic factors are equal is valid only for
four-cluster model calculations of
$^{13}$N -$^{13}$C and for ground states of $^{17}$F - $^{17}$O. 
However, the deviation
between ${\cal R}_{s.p.}$, obtained with this assumption, 
and microscopic calculations in most cases
is not strong, being of the the same order as   ${\cal R}_0/{\cal R}_{MCM}$.  

The predictions from MCM can be used to calculate proton ANCs using
experimentally determined neutron ANCs
and vice versa. As an example, let us calculate ANCs for $^8$B  
from experimentally determined values $C^2_{1\frac{3}{2}}(^8$Li) =
0.384 $\pm$ 0.038 fm$^{-1}$ and $C^2_{1\frac{1}{2}}(^8$Li) =
0.048 $\pm$ 0.006 fm$^{-1}$ from Ref. \cite{Livius}. With
${\cal R}_{\frac{3}{2}}$ and ${\cal R}_{\frac{1}{2}}$ values from
Table I we get that $C^2_{1}(^8$B) is 0.460 $\pm$ 0.048 fm$^{-1}$
for V2 and  0.471 $\pm$ 0.048 fm$^{-1}$ for MN. This values give
 the astrophysical $S$-factor of the $^7$Be(p,$\gamma)^8$B reaction
at zero energy
$S_{17}(0)$ =  17.8 $\pm$ 1.7 eV$\cdot$b for  V2 and 18.2 $\pm$ 1.8 eV$\cdot$b 
for the MN. The difference between these two
calculations is only 2$\%$.

Finally, if theoretical predictions for the ratio between mirror ANCs 
are not available, simultaneous consideration of
analytical formula (\ref{rn}) and of single-particle estimate ${\cal R}_{s.p.}$
can be used. Based on our calculations, the average between 
these values may be a reasonably good
approximation if the core is not strongly deformed. Strong core polarization
effects can increase this ratio. The largest increase, 
calculated in the present paper, is 12$\%$.

\section*{Acknowledgements}
N.K.T. is grateful to Professors R.C. Johnson and I.J. Thompson for
fruitful discussions and useful comments concerning this manuscript. 
Support from the UK EPSRC via grant GR/T28577 is gratefully acknowledged.

\end{document}